\documentstyle[12pt,psfig]{article}
\def\reference{\parskip 0pt\par\noindent\hangindent 0.5 truecm}
\begin{document}
\newcommand{\fnhi}{\mbox{$f(N_{\rm HI})$}}
\newcommand{\fn}{\mbox{$f(N)$}} \newcommand{\nhi}{\mbox{$N_{\rm HI}$}}
\newcommand{\hi}{H~{\sc i}} \newcommand{\dla}{DL$\alpha$}
\newcommand{\mhi}{\mbox{$M_{\rm HI}$}}
\newcommand{\mhis}{\mbox{$M^*_{\rm HI}$}}
\newcommand{\msol}{\mbox{${\rm M}_\odot$}}
\newcommand{\phis}{\mbox{$\phi^*$}}
\newcommand{\ohi}{\mbox{$\Omega_{\rm HI}$}}
\newcommand{\rhi}{\mbox{$\rho_{\rm HI}$}}
\newcommand{\hmpc}{\mbox{$h_{100}^{-1}\, \rm Mpc$}}
\newcommand{\ihmpc}{\mbox{$h_{100}\, \rm Mpc^{-1}$}}
\newcommand{\hmpcc}{\mbox{$h_{100}^{-3}\, \rm Mpc^{3}$}}
\newcommand{\hmpmpcc}{\mbox{$h_{100}\, \rm M_\odot Mpc^{-3}$}}
\newcommand{\hgpcmc}{\mbox{$h_{100}\, \rm g \,cm^{-3}$}}
\newcommand{\ihmpcc}{\mbox{$h_{100}^{3}\, \rm Mpc^{-3}$}}
\newcommand{\kms}{\mbox{$\rm km\, s^{-1}$}}
\newcommand{\magsq}{\mbox{$\rm mag\, arcsec^{-2}$}}
\newcommand{\icmsq}{\mbox{$\rm cm^{-2}$}}
%
%
\title{The \hi\ Column Density Distribution Function at $z=0$: the
  Connection to Damped Ly$\alpha$ Statistics}
%


\author{Martin A. Zwaan $^{1}$ \and Marc A. W. Verheijen $^{2}$ \and
  Frank H. Briggs $^{1}$
} 

\date{} \maketitle

{\center $^1$ Kapteyn Astronomical Institute, P.O. Box 800, 9700 AV
  Groningen, The Netherlands\\zwaan, fbriggs@astro.rug.nl\\[3mm] $^2$
  National Radio Astronomy Observatory, P.O. Box 0, Socorro, NM 87801,
  U.S.A.\\mverheij@aoc.nrao.edu\\[3mm] }

%
\begin{abstract} 
  We present a measurement of the \hi\ column density distribution
  function [\fnhi] at the present epoch for column densities
  $>10^{20}~\icmsq$.  These high column densities compare to those
  measured in damped Ly$\alpha$ lines seen in absorption against
  background quasars.  Although observationally rare, it appears that
  the bulk of the neutral gas in the Universe is associated with these
  damped Ly$\alpha$ systems.  In order to obtain a good anchor point
  at $z=0$ we determine \fnhi\ in the local Universe by using 21cm
  synthesis observations of a complete sample of spiral galaxies. We
  show that \fnhi\ for damped Ly$\alpha$ systems has changed
  significantly from high $z$ to the present and that change is
  greatest for the highest column densities.  The measurements
  indicate that low surface brightness galaxies make a minor
  contribution to the cross section for \hi, especially for
  $\nhi>10^{21}~\icmsq$.
 \end{abstract}

\bigskip

%
%

\section{Introduction}
High column density absorbers seen in the spectra of background QSOs
are referred to as Damped Ly$\alpha$ (\dla) systems if the observed
\hi\ column density exceeds the value of $\nhi=2\times
10^{20}~\icmsq$.  The \dla\ absorption lines are on the square-root
part of the curve of growth where damping wings dominate the profile
and column densities can be determined accurately by fitting the line
profiles. Wolfe (1995) argues that these systems at high redshifts are
gas-rich disks in the process of contracting to present-day spiral
galaxies.  This idea is supported by the fact that the characteristic
velocity profiles of metal lines and Lyman series lines in \dla\ 
systems are similar to those of sightlines through spiral galaxies at
$z=0$.  More recently, detailed modeling of \dla\ absorption profiles
by Prochaska \& Wolfe (1998) has shown that the \dla\ systems are
consistent with rapidly rotating, thick disks.  Note however that
alternative models, like protogalactic clumps coalescing into dark
matter halos (Haehnelt et al.  1997, Khersonsky \& Turnshek 1996), can
also explain the kinematics. The cosmological mass density of neutral
gas in \dla\ systems at high redshift is comparable to the mass
density of luminous matter in galaxies at $z=0$ (e.g.  Lanzetta et al.
1995).

One of the best known statistical results of the study of QSO
absorption line systems is the column density distribution function
(CDDF) of neutral hydrogen.  The function describes the chance of
finding an absorber of a certain \hi\ column density along a random
line of sight per unit distance.  An observational fact from high-$z$
Ly$\alpha$ studies is that the differential CDDF [\fnhi] can be
described by a single power law of the form $\fnhi \propto
\nhi^{\alpha}$, where $\alpha\approx -1.5$ over ten orders of
magnitude in column density (e.g.  Tytler 1987, Hu et al.  1995) from
$10^{12}~\icmsq$ (Ly$\alpha$ forest) to $10^{22}~\icmsq$ (\dla).

An integration over the distribution function gives the total
cosmological neutral gas density as a function of redshift.  The \hi\ 
gas density relates to \fnhi\ as $\ohi \propto \int_{N_1}^{N_2} \nhi
\fnhi d\nhi$ and it is readily seen that $\ohi (\nhi) \propto
N_2^{0.5}$ if $\alpha=-1.5$ and $N_2 \gg N_1$.  This implies that
although the high column density systems are observationally rare,
they contain the bulk of the neutral gas mass in the Universe.
Because so few \dla\ systems are known ($\approx 80$), the
uncertainties on \ohi\ and the CDDF for high column densities are
large, especially if the measurements are split up into different
redshift bins.  But following the CDDF as a function of redshift is
certainly very important in constraining models of complicated
physical processes like star formation or gas feedback to the
interstellar medium.

There are several reasons why the determination of \fnhi\ at the
present epoch is difficult.  Due to the expansion of the Universe the
expected number of absorbers along a line of sight decreases with
decreasing redshift, the Ly$\alpha$ line is not observable from the
ground for redshifts smaller than 1.65, and starlight and dust in the
absorbing foreground galaxies hinder the identification of the
background quasars.  Gravitational lensing may also play a role as it
can bring faint quasars into the sample which otherwise would not have
been selected (e.g. Smette et al. 1997).

At the present epoch the largest repositories of neutral gas are
clearly galaxies. No instance of a free-floating \hi\ cloud not
confined to the gravitational potential of a galaxy has yet been
identified. It is therefore justified to use our knowledge of the
local galaxy population to estimate the shape and normalization of the
CDDF.

\section{How to determine \fnhi\ at $z=0$?}
\begin{figure}
\begin{center}
  \centerline{\psfig{file=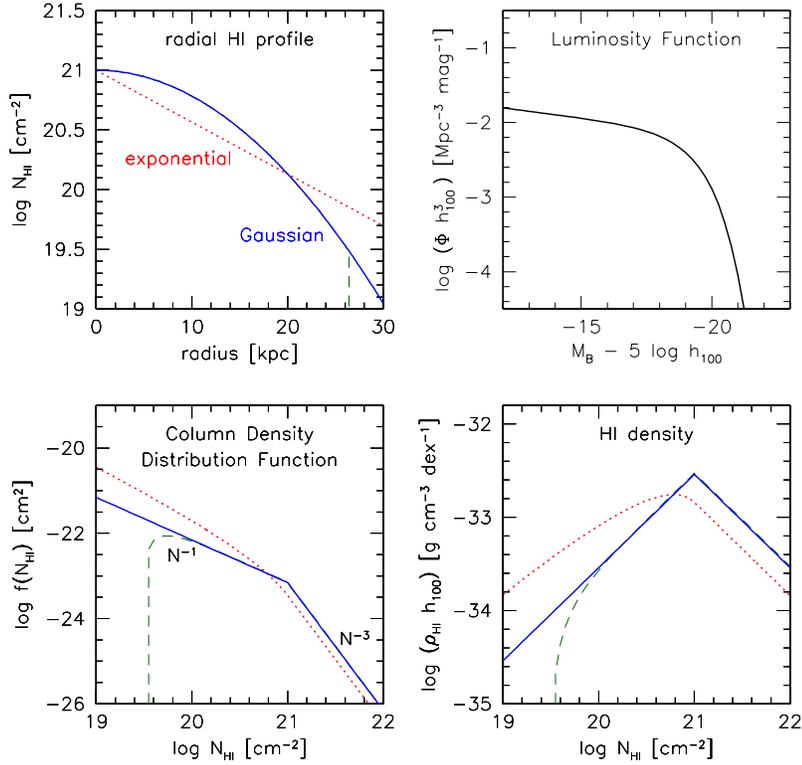,height=11cm}}
 \caption{Illustration of simple models for the CDDF. 
   {\em Upper left\/}: Gaussian and exponential models for the radial
   distribution of neutral gas density in galactic disks.  {\em Upper
     right\/}: Schechter function describing the local luminosity
   function.  {\em Lower left\/}: Resulting CDDFs for the Gaussian and
   exponential radial profiles.  {\em Lower right\/}: Integral neutral
   gas density in the local Universe as a function of column density
   for both models.  The effect of disk truncation is indicated by the
   dashed line.}
 \label{modelfn.fig}
 \end{center}
 \end{figure}

A simple but illustrative and instructive method is to take the
analytical approach.  This is illustrated in Figure~\ref{modelfn.fig}.
Here we represent the radial distribution of the neutral hydrogen gas
in galaxies by both an exponential and a Gaussian model.  The
differential cross sectional area of an inclined ring with a column
density in the range $N$ to $N + dN$ is given by $d \Sigma (N,i) = 2
\pi r(N) dr \cos i$, where $r(N)$ is the radius at which a column
density $N$ is seen, and $i$ is the inclination of the ring.  We
assume that the luminosity function $\phi(M)$ of the local galaxy
population can be described by a Schechter function as indicated in
the upper right panel of figure~1.  The local $f(N)$ can be derived
from $\phi(M)$ and the area function $d\Sigma(N)$ by taking the
integral
\begin{equation}
   f(N) = \frac{c}{H_0} \frac{\int_{M_{\rm min}}^{M_{\rm max}} \phi(M)
     \langle d\Sigma (N) \rangle_i \, dM}{dN},
\end{equation}   
where the subscript $i$ indicates an average over all inclinations.
To evaluate this integral, the area function, or more generally the
radial \hi\ distribution, needs to be related to $M$.  Here we adopt
the relation $\log \mhi=A+B M_B$ (following Rao \& Briggs 1993) and
assume that the central gas surface density in disks is not dependent
on morphological type or luminosity.  The resulting $f(N)$ for both
models is shown in the lower left panel.  The integral \hi\ gas
density in \hgpcmc\ as a function of column density is shown in the
lower right panel.  This function can be calculated with $\rhi(N) =
m_{\rm H} N \frac{H_0}{c}f(N) dN$, where $m_{\rm H}$ is the mass of
the hydrogen atom.

The Gaussian models yield a CDDF of the form $\fn\propto N^{\alpha}$,
where $\alpha=-1$ for $N$ smaller than the maximum column density seen
in a face-on disk ($N_{\rm max}$) and $\alpha=-3$ for the higher
values of $N$.  The exponential model gives a smoother function.  The
logarithmic slope is approximately $-1.2$ around $N=10^{20}~\icmsq$,
slowly changing to $-3$ at higher column densities.  In fact, it was
shown already by Milgrom (1988) that $\fn \propto N^{-3}$ for
$N>N_{\rm max}$ for any radial surface density distribution.  The
lower right panel clearly illustrates that an overwhelming part of the
total \hi\ mass in the local Universe is associated with column
densities close to $N_{\rm max}$.

In addition to these simple models we also show the effect of disk
truncation on the CDDF.  The thin dashed line illustrates a Gaussian
disk truncated at $\nhi=10^{19.5}~\icmsq$, the level below which
photo-ionization by the extragalactic UV-background is normally
assumed to be important (e.g. Corbelli \& Salpeter 1993, Maloney
1993). It appears that this truncation only seriously affects the CDDF
below $\nhi=10^{19.5}~\icmsq$. No significant changes occur at higher
column densities.

 A more reliable method than this analytical approach is to determine
 \fnhi\ by using observed \hi\ distributions.  21cm maps of nearby
 galaxies routinely reach sensitivity limits comparable to column
 densities that typify \dla\ absorbers.  It therefore seems natural to
 calculate $\fnhi(z=0)$, simply by adding cross sectional areas as a
 function of \nhi\ for a large sample of galaxies for which 21cm
 synthesis observations are available.  However, this approach is
 complicated by the fact that there is an enormous variation in
 sensitivity and angular resolution of the 21cm maps, and the problem
 of choosing a fair and complete sample of galaxies.  Most galaxies
 that have been studied extensively in the 21cm line were selected on
 having either a large \hi\ diameter, so that the rotation curve can
 be sampled out to large galactocentric radii, or on having
 peculiarities such as polar rings or warps.  Thus, most samples for
 which 21cm synthesis data exist are not representative of the galaxy
 population of the local Universe and would likely be biased against
 dwarf and low surface brightness galaxies.

\section{The Ursa Major Cluster}
The Ursa Major cluster of galaxies, studied extensively by Verheijen
(1997), forms an ideal sample for an unbiased study of the column
density distribution function.  The Ursa Major cluster, at a distance
of 15 Mpc, is different in many respects from famous, equally distant
clusters like Virgo and Fornax.  Ursa Major's member galaxies show no
concentration towards a central condensation and their velocity
dispersion is exceptionally low, approximately 150 km/s.  The
estimated crossing time is half a Hubble time and hence the galaxies
do not seem to be seriously affected by tidal interactions.  In
addition to this, there is a predominance of late type galaxies and
the morphological mix of galaxies is indistinguishable from that in
the field.  This combination of properties implies that the Ursa Major
cluster is in fact an overdensity of galaxies and not comparable to
classical clusters of galaxies. This justifies the use of the Ursa
Major cluster for the study of the shape of the CDDF of neutral
hydrogen in the local Universe.

The Ursa Major cluster as defined by Tully et al. (1996) comprises a
volume of 80 Mpc$^3$, within which 80 galaxies are identified to date.
For a complete sample of 62 galaxies intrinsically brighter than the
Small Magellanic Cloud ($M_B=-16.5^{\rm m}$) 21cm synthesis
observations have been performed with the WSRT\footnote{The WSRT is
  operated by the Netherlands Foundation for Research in Astronomy
  (NFRA/ASTRON), with financial support by the Netherlands
  Organization for Scientific Research (NWO)}.  \hi\ has been detected
by the WSRT in 49 galaxies.  Details on observations and data reduction are
described in Verheijen (1997).

An obvious advantage of using the UMa sample for this study is that
all the member galaxies are at approximately the same distance.
Therefore, the spatial resolution of the synthesis observations are
constant for the whole sample. This simplifies the problem of
assessing the influence of resolution on the determination of the CDDF
and the comparison with the CDDF at high redshift.

The shape of the column density distribution function is determined by
counting in each \hi\ map the number of pixels per logarithmic bin of
0.1 dex in column density.  The solid angle covered by pixels of a
certain column density is then determined by multiplying the number of
pixels with the angular pixel size which varies slightly from galaxy
to galaxy.

The disadvantage of using a galaxy sample taken from a clear cosmic
overdensity is that the CDDF is not automatically normalized.  If we
would naively assume that the Ursa Major cluster is a representative
part of the nearby Universe, we would overestimate the normalization
of the CDDF by roughly a factor of 12.  This factor is obtained by
comparing the \hi\ mass function of the cluster with that of the field
galaxy population (Zwaan et al.  1997).  The shape of the Ursa Major
mass function is indistinguishable from that of the field, but the
normalization, $\theta^*$, is larger by a factor of $\sim 12$.
Ideally, one would use a sample of galaxies with well understood
selection criteria so that the normalization would occur
automatically.  Unfortunately, there are no such samples available for
which \hi\ synthesis observations with sufficient angular resolution
have been performed.  The HIPASS survey, a blind 21cm survey of the
whole southern sky, will eventually yield a suitable galaxy sample for
this purpose, if a representative subsample is followed up with the
ATCA to obtain high spatial resolution maps.

There are several methods for normalizing the UMa CDDF.  By assuming a
local luminosity function (LF) or \hi\ mass function (HIMF), each
galaxy could be given a weight according to its absolute magnitude or
\hi\ mass.  However, this method introduces extra uncertainty in the
derived CDDF, due to uncertainties in the exact shape and
normalization of the LF and the HIMF.  Our preferred method of
normalizing the CDDF is to scale the complete function, not the
individual contributors to it.  This can be achieved by scaling the
integral \hi\ mass density that is contained under the CDDF:
 \begin{equation} \rho_{\rm HI} = \int_{N_{\rm
       min}}^{N_{\rm max}} m_{\rm H} N \frac{H_0}{c} f(N) dN.
 \end{equation}
 By means of a blind 21cm survey Zwaan et al.  (1997) determined
 $\rho_{\rm HI}= 5.8 \times 10^7 \hmpmpcc$, a result that is in
 excellent agreement with earlier estimates based on optically
 selected galaxies. Note that dependencies on $H_0$ disappear in the
 final specification of the CDDF.

\section{The Column Density Distribution Function}
\begin{figure}
 \begin{center}
   \centerline{\psfig{file=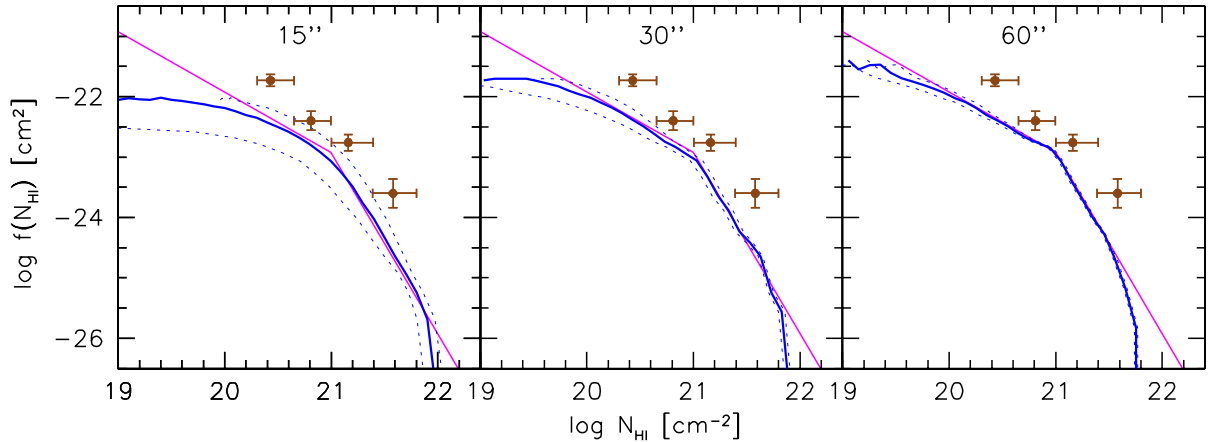,height=7.3cm}} 
 \caption{The column density distribution function at $z=0$.  From left
   to right the
   function is shown for three different resolutions of the \hi\ maps:
   $15''$, $30''$, and $60''$.  The thick
   solid line is the CDDF as measured in the Ursa Major cluster.  The
   dashed lines indicate the average $1\sigma$ uncertainties in the
   column density determinations.  For reference, a Gaussian model
   with $N_{\rm max}=10^{21}~\icmsq$ is also shown as a thin solid
   line.  The points indicate the \hi\ CDDF for damped Ly$\alpha$
   systems at high redshift ($\langle z\rangle\approx 2.5$) taken from
   Storrie-Lombardi et al. (1997).  Note the strong evolution,
   especially at the highest column densities. }
 \label{umafn.fig}
 \end{center}
 \end{figure}

Figure~\ref{umafn.fig} shows the CDDF determined from the 21cm
observations of the Ursa Major sample.  From left to right the
function is shown for three different resolutions of the \hi\ maps:
$15''$, $30''$, and $60''$.  The solid line is the determined CDDF;
the dashed lines indicate the quality of the measured column
densities.  Each pixel in the \hi\ maps has an estimate of the signal
to noise level assigned to it.  In the determination of the CDDF we
calculated an average S/N level for each bin in column density by
averaging the S/N ratios for the individual pixels.  The dashed lines
show the average $1\sigma$ errors on the column densities and should
be interpreted as horizontal errorbars.  Nonetheless, they clearly
overestimate the real uncertainties on the CDDF as many pixels are
used in each bin (2500 independent beams for full resolution).  The
lines merely serve as an indicator of the quality of the measurements
at each resolution.  The thin solid line represents the CDDF for a
Gaussian model, where $\fnhi \propto \nhi^{-1}$ for
$\nhi<10^{21}~\icmsq$ and $\fnhi \propto \nhi^{-3}$ for higher column
densities.

 When comparing the CDDFs at different resolution, it appears that the
 highest resolution maps yield the smoothest CDDF.  This occurs
 because there the measurements of column density have the lowest S/N
 ratios.  The low resolution (but high S/N) CDDF is in excellent
 agreement with the Gaussian model for $10^{20}<N<10^{21.5}$, but for
 higher column densities, the measured curve drops below the model
 since high column density peaks are smeared away.  Going to higher
 resolutions leads to better agreement between the measured curve and
 the model for the highest \nhi, and at $15''$ resolution the CDDF
 follows the $\nhi^{-3}$ distribution up to $10^{21.9}~\icmsq$.

 Besides beam smearing two other effects can cause a deviation from
 the $\nhi^{-3}$ function.  Firstly, the calculations assume that the
 gaseous disks are infinitely thin.  Observations show that the bulk
 of the \hi\ indeed resides in a thin layer with axis ratio $< 0.1$
 (Rupen 1991).  The thin disk approximation is therefore valid for
 moderately inclined disks.  However, the highest column densities in
 the models arise in highly inclined thin disks.  A small degree of
 puffiness will prevent these high column densities from being
 observed.  The second effect is \hi\ self absorption. The theoretical
 calculation of the CDDF is based on the assumption that the optical
 depth of the neutral gas layer is negligible.  Column densities much
 higher than the maximal column density in a face-on galaxy can only
 be seen in a highly inclined disk when the gas is optically thin.  It
 is remarkable that the full resolution CDDF follows the $f(N) \propto
 \nhi^{-3}$ line up to $\nhi=10^{21.9}$, well above the value where
 \hi\ self absorption is normally assumed to set in.  For example,
 Dickey \& Lockman (1990) calculate that an \hi\ cloud with $T=50 \rm
 K$ and an FWHM velocity dispersion of $10~\kms$ becomes optically
 thick ($\tau=1$) at column densities $N=10^{21}~\icmsq$.

 Also shown in Figure~\ref{umafn.fig} are the measurements of \fnhi\ 
 at high redshifts as determined by Storrie-Lombardi et al.  (1997).
 We choose not to split up their high-$z$ sample in different redshift
 bins in order to get reasonable signal to noise.  The median redshift
 of the total \dla\ sample is $z=2.5$.  A value of $q_0=0.5$ has been
 used here.  Lower values of $q_0$ would not significantly change the
 slope of \fnhi\ but would decrease the normalization by approximately
 a factor of 2.  Strong redshift evolution of the CDDF from $z=2.5$ to
 the present is apparent.  The intersection cross-section for \hi\ 
 column density $<10^{21.2}~\icmsq$ has decreased by a factor of 6
 (factor 3 for $q_0=0$) from $z=2.5$ to $z=0$.  Higher column
 densities show a larger decrease, the evolution accounting for a factor
 of 10 (5 for $q_0=0$).  Lanzetta et al (1995) report still stronger
 evolution of the higher column densities for higher redshift,
 although the highest column densities suffer from small number
 statistics and the effect is hardly seen by Storrie-Lombardi et al.
 (1997).  The strong evolution of the higher column densities can be
 understood if gas consumption by star formation occurs most rapidly
 in regions of high neutral gas density (Kennicutt et al. 1994).

 Rao \& Briggs (1993) evaluated the CDDF at the present epoch by
 analyzing Arecibo observations of a sample of 27 galaxies with
 optical diameters in excess of $7'$.  Double-Gaussian fits to the
 observed radial \hi\ distribution were used to calculate \fnhi. The
 disadvantage of this method is that the Gaussian fits automatically
 introduce the $\nhi^{-1}$ for low \nhi\ and $\nhi^{-3}$ for high
 \nhi. In the present study no modeling has been applied. The location
 of the change of the slope and the normalization are in excellent
 agreement between Rao \& Briggs' work and the Ursa Major
 determination.

\section{Contribution of Low Surface Brightness Galaxies}
It has been argued in the literature that low surface brightness (LSB)
galaxies might contribute a considerable \hi\ cross section.  In
particular, Linder (1998) explores a scenario in which the outskirts
of galaxies are responsible for most of the cross section for low
column density neutral gas ($\nhi < 10^{20.3}~\icmsq$).  She concludes
that Ly$\alpha$ absorber counts at low redshifts can be explained if
LSB galaxies of moderate absolute luminosity with extended low density
gas disks are included in the analysis.  Contrary to this view, Chen
et al.  (1998) claim that extended disks of luminous galaxies can
account for most of the observed Ly$\alpha$ lines below
$\nhi=10^{20.3}~\icmsq$.  The contribution of dwarf and LSB galaxies
to the cross section for high column density \hi\ is also unclear.
For instance, Rao \& Turnshek (1998) show that there are no luminous
spiral galaxies in the vicinity of the quasar OI~363 in which spectrum
they identify two low-$z$ \dla\ systems.

Here we evaluate the contribution of LSB galaxies to the cross section
for high column density gas at $z=0$.  First we have to address the
problem of completeness.  The Ursa Major sample is essentially a
magnitude limited sample.  Selection effects against LSB galaxies are
therefore to be expected.  Tully \& Verheijen (1997) discuss the
completeness of the sample by plotting the observed central surface
brightness against the exponential disk scale length.  Theoretical
approximations of the visibility limits seem to describe the
boundaries of the observed sample satisfactorily.  We apply the same
visibility limits to the \hi\ selected galaxy sample of Zwaan et al.
(1997) to estimate what fraction of the \hi\ mass density in the Ursa
Major cluster could be missed in the present study.  It appears that
galaxies below the optical detection limits of our Ursa Major sample
probably contain 10\% of the total \hi\ density of the cluster.  Not
surprisingly, most of this missed \hi\ density resides in LSB
galaxies.  Following Tully \& Verheijen (1997), the separation between
LSB and HSB galaxies is made at an extrapolated central surface
brightness of $18.5~\magsq$ in the $K'$-band, which roughly compares
to $22.0~\magsq$ in the $B$-band.

\begin{figure}
 \begin{center}
   \centerline{\psfig{file=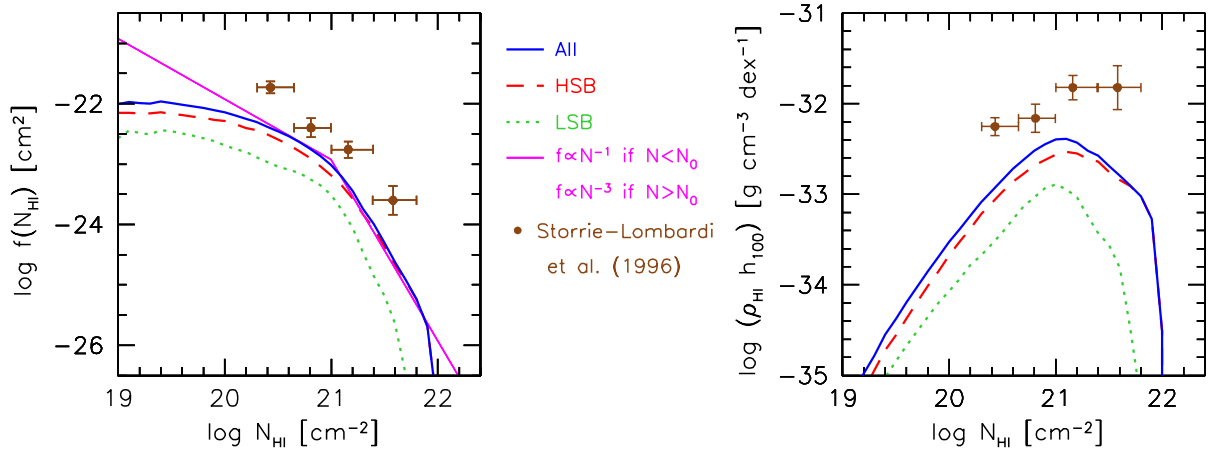,height=7.3cm}}
 \caption{The column density distribution function ({\em left\/}) and
   \hi\ mass density ({\em right\/}) for high and low surface
   brightness galaxies.  High surface brightness galaxies dominate the
   \hi\ cross section, especially for column densities
   $>10^{21.3}~\icmsq$.}
 \label{lsbs.fig}
 \end{center}
 \end{figure}

Figure~\ref{lsbs.fig} illustrates the contribution of LSB galaxies to
the cross section for high column density gas, relative to the total
galaxy population.  We have corrected for the incompleteness by adding
extra cross section for the LSB galaxies, equally over all column
densities, in such a way that the mass density in these galaxies
increases by an amount equal to 10\% of the total \hi\ density.  The
left panel shows the CDDF, the right panel shows the cosmological mass
density of \hi\ as a function of column density.  The full resolution
data are used.  LSB galaxies do not make a significant contribution to
the cross section for column densities higher than
$\nhi=10^{21.3}~\icmsq$.  Below that value they are responsible for
approximately 25\% of the cross section.  The right panel shows that
LSB galaxies make a minor contribution to the local neutral gas
density, a conclusion very much in concordance with the results of
Briggs (1997), Zwaan et al.  (1997), Sprayberry (1998) and C\^ot\'e et
al.  (1998).

\section{Conclusions and Discussion}
We have used the present knowledge of the nearby galaxy population to
estimate the \hi\ column density distribution function at $z=0$.  It
is shown that \fnhi\ undergoes strong redshift evolution from
$z\sim2.5$ to the present, especially at the high column densities.
The observed evolution in \fnhi\ critically depends on whether the
census of \hi\ in the local Universe is complete.  Surveys in \hi\ and
the optical indicate that the density of visible light and neutral gas
is dominated by luminous, high surface brightness galaxies.  The \hi\ 
surveys routinely reach column density limits much lower than what is
required to detect the $z=0$ counterparts of \dla\ systems.  Since
\hi\ mass functions published to date typically lose sensitivity below
$\mhi=10^7 \msol$, the region of parameter space still open to hide a
large amount of high column density gas is that of low \hi\ masses.
Observations to measure the space density of these small \hi\ masses
(\hi\ clouds and extreme LSB dwarf galaxies) and to evaluate to what
extent they contribute to the \hi\ density and the CDDF of the local
Universe are important next steps.






\section*{References}
\reference Briggs, F.H. 1997, ApJ, 484, 618
\reference Chen, H.-W. Lanzetta, K.M., Webb, J.K., \& Barcons, X. 1998,
        ApJ, 498, 77
\reference Corbelli, E.\& Salpeter, E. E. 1993, ApJ, 419, 104
\reference C\^ot\'e, S, Broadhurst, T., Loveday, J., \& Kolind, S. 1998, astro-ph/9810470
\reference Dickey, J.M. \& Lockman, F.J. 1990, ARA\&A, 28, 215   
\reference Haehnelt, M.G., Steinmetz, M., \& Rauch, M. 1998, ApJ 
        495, 647
\reference Hu, E.M., Kim, T.-S., Cowie, L.L., Songaila, A., \& Rauch,
        M. 1995, AJ, 110, 1526
\reference Kennicutt, R.C. 1998, ApJ, 498, 541
\reference Khersonsky, V.K. \& Turnshek, D.A. 1996, ApJ, 471, 657
\reference Lanzetta,  K.M., Wolfe, A.M., \& Turnshek, D.A
        1995, ApJ, 440, 435
\reference Linder, S.M. 1998, ApJ, 495, 637   
\reference Maloney, P. 1993, ApJ, 414, 41
\reference Milgrom, M. 1988, A\&A, 202, L9
\reference Prochaska, J.X. \& Wolfe, A.M. 1997, ApJ, 487, 73
\reference Rao, S., \& Briggs, F.  H.  1993, ApJ, 419, 515
\reference Rupen, M.P. 1991, AJ, 102, 48
\reference Smette, A., Claeskens, J.F., \& Surdej, J. 1997, NewA, 2, 53
\reference Sprayberry, D. 1998 these proceedings
\reference Storrie-Lombardi, L.J., Irwin, M.J., \& McMahon, R.G.
        1996, MNRAS, 282, 1330 
\reference Tully, R.B., Verheijen, M.A.W., Pierce, M.J., Huang,
        J.S., \& Wainscoat, R.J. 1996, AJ, 112, 2471
\reference Tully, R.B. \& Verheijen, M.A.W. 1997, ApJ, 484, 145 
\reference Tytler, D. 1987, ApJ, 321, 49
\reference Verheijen, M.A.W. 1997, Ph.D. thesis, Univ. Groningen
\reference Wolfe 1995, in QSO Absorption Lines, ed. G.~Meylan
\reference Zwaan, M.A., Briggs, F.H., Sprayberry, D., \& Sorar, E.
        1997, ApJ, 490, 173






\end{document}